\documentclass{epl}
\usepackage{amsmath}
\usepackage{amssymb}

\title{Plasmon assisted transport through disordered array of quantum wires}
\author{A. L. Chudnovskiy}
\institute{
I. Institut f\"ur Theoretische Physik, Universit\"at Hamburg,
Jungiusstr. 9, 20355 Hamburg, Germany
}

\begin{document}

\maketitle

\begin{abstract}
Phononless plasmon assisted transport through a
long disordered array of finite length quantum wires is
investigated analytically.  Generically strong electron plasmon interaction
in quantum wires results  in  a qualitative change of the temperature dependence
of thermally activated resistance in comparison to phonon assisted transport.
At high temperatures, the thermally activated resistance is determined by the
Luttinger liquid interaction parameter of the wires.
\end{abstract}

Hopping transport in a quasi-one-dimensional system formed by a parallel arrangement
of conducting wires is of much relevance to a number of experimental setups, including
quantum wire arrays in heterojunctions \cite{Mani}, carbon nanotube films \cite{deHeer},
atomic wires on silicon surface \cite{Himpsel}, and stripe phases \cite{Fogler}.
At finite length of constituent wires, such systems represent particular
examples of granular arrays, where a one-dimensional wire  plays the role of a grain.
Considered as a granular array, the array of parallel quantum wires is rather peculiar
because of the very long charge relaxation time in a one-dimensional wire.
Due to this peculiarity, the theoretical description of thermally activated transport
in arrays of long quantum wires requires  taking into account the charge
dynamics and treatment of the interactions beyond the capacitive model adopted in recent
theoretical investigations of transport through disordered granular arrays
\cite{AGK}.

In this letter we show that charge-density fluctuations
(plasmons) in the array can act as the agent promoting thermally activated transport,
thus providing the possibility for phononless inelastic transport.
As the result of generically strong plasmon-electron coupling in
a quantum wire, the features of plasmon and phonon assisted transport are
qualitatively  different.
We provide a qualitative explanation
of plasmon assisted transport, identify the transport regimes, where the features of
plasmon and phonon assisted transport are either similar or substantially different, and
derive analytic expressions for the temperature dependence of the thermally activated
resistance for a special model of a strongly correlated disordered array of quantum
wires.

The model we formulate below is special, because it combines two seemingly incompatible
features: {\bf i)} it is strongly disordered for single electron transport;
{\bf ii)} it is much weaker disordered for propagation of plasmons.

\begin{figure}
\onefigure[width=10cm,height=5cm,angle=0]{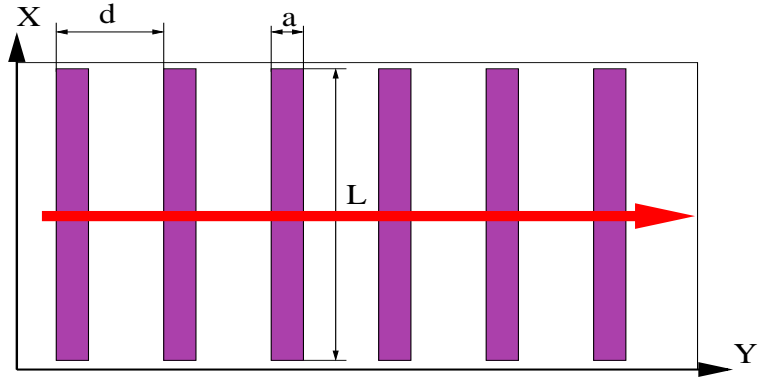}
\caption{Geometry of the model. The arrow shows the direction of the current. }
\label{figsetup}
\end{figure}

Consider a one dimensional array of parallel  identical quantum wires of length
$L$ and diameter $a$ placed regularly with the interwire distance $d$, $L\gg d\gg a$
(see Fig. \ref{figsetup}).
We investigate transport in the
direction perpendicular to the wires. The spectrum of low-energy plasmons in a single
isolated wire is equidistant with energies
$E_{i,n}=\frac{\pi v_i}{L}n$, where $L$ is the length of the wire, and $v_i$ is the plasmon
velocity along the wire $i$.
For identical wires at regular positions, the intra- and
interwire interactions between the charge density fluctuations do not change along the array.
Then each localized plasmon level broadens into a plasmon band
with truly continuous spectrum, quite analogously to the formation of electronic bands
in the tight binding model. The role of the hopping in the tight-binding model
for plasmons is played by the matrix element of the charge-density interactions between the
neighbor wires.
The formation of plasmon bands is reflected by the dependence of
the plasmon velocity along each wire on the wave vector $p$ along the array (that is,
perpendicular to the wires)
$v_0\rightarrow u(p)$. The particular form of plasmon dispersion depends on details of the
interwire interactions, yet the function $u(p)$ should be periodic with
a period of one Brillouin zone. That is why we choose the specific form
\begin{equation}
u_p=v_0+v_1\cos(\pi p), \ -1<p\leq 1.
\label{u_p}
\end{equation}
The chosen form of dispersion can be considered as the first two terms of the Fourier
expansion of some general dispersion law. The plasmon energy within a band centered around
the level $n$ is given by $\epsilon_{n}(p)=\frac{\pi u_p}{L}n$.

Furthermore, if the potential energy barriers between the wires are high,
the electronic bands for the motion along the array (in Y-direction in Fig. \ref{figsetup})
are very narrow with the width
$t\approx\frac{\hbar^2\lambda^3d}{2m_*(a\lambda+1)}e^{-\lambda d}$,
where $\lambda$ describes  the decay of the single particle wave function inside the barrier,
$m_*$ is the effective mass of an electron.
Now let us introduce the disorder as a random height of the energy barriers between the neighbor
wires. Such a disorder induces fluctuations of the decay parameter $\lambda$ thus rendering $t$
random. As the result, the single particle wave functions become localized. At
$\langle\delta\lambda^2\rangle\approx\frac{\overline{\lambda}^4d^2}{8\pi}
e^{-2\overline{\lambda}d}$,
the localization length of the single particle wave
function is of order $d$, that is, an electron is localized within a single wire.

The resistance of a long disordered one-dimensional array is determined by so-called
breaks, the junctions between two neighboring  wires with exponentially high resistance
\cite{RR}. Let us denote the energy cost to transfer an electron over the break as  $E_a$.
To facilitate the transport over the break, the energy $E_a$ should be
borrowed by absorption of a bosonic excitation. Now we show that
due to the transfer of energy between charge density excitations in
different wires, a bosonic bath with continuous energy spectrum can be created at
the break. Then any energy deficit $E_a$ can be matched by absorption of a
plasmonic excitation, which constitutes the microscopic mechanism of plasmon
assisted transport \cite{Tigran}.
For two isolated wires forming the break, the matching condition
$E_{i,n}=E_a$ cannot be satisfied for arbitrary $E_a$ because of discreteness of $E_{i,n}$.
However, due to the charge-density interwire interaction
$H_{int}=\sum_i\sum_{n,n'}V^{nn'}_{i,i+1}\hat{\rho}_i(n)\hat{\rho}_{i+1}(n')$,
where $n, n'$ mark the plasmon modes, the energy can
be transferred between excitations localized in different wires.
Treating the interwire interaction
perturbatively, we deduce, that the transfer of  plasmon
energy from the nearest neighbor of the wire $i$ leads to splitting of the plasmon
level $E_{i,n}$ into $M$ sublevels with energies
$\epsilon_{i,n}^{l}=E_{i,n}-\frac{|V^{nl}_{i,i+1}|^2}{E_{i+1,l}-E_{i,n}}$,
where $l=\overline{1,M}$ runs over the plasmon levels of the wire $i+1$, and
$M$ denotes the number of levels that have a
nonvanishing interaction strength with the level $(i,n)$.
Each split energy sublevel corresponds to a complex plasmon excitation,
which is a superposition of plasmons localized in the two wires.
Note that all the split sublevels are situated in
the energy window $\frac{|V|^2}{\delta_1}$ around $E_{i,n}$,
where $V$ is a typical value of the interwire interaction,
and $\delta_1$ is the plasmon interlevel distance in a
single wire,  $|V|\ll\delta_1$.
Now we take into account that each energy level of the wire $i+1$ is in turn  split
due to the interaction with the next-nearest wire $i+2$. Repeating the previous calculation
with $E_{i+1,l}$ substituted by the energy of a split level  $\epsilon_{i+1,l}^{m}$ we
conclude that the energy interval $|V|^2/\delta_1$ around $E_{i,n}$ is now filled with
$M^2$ levels. By recursion we obtain that the transfer of energy from $n$ neighbors of
the break results in $M^n$ plasmon sublevels within the energy interval
$|V|^2/\delta_1$ around $E_{i,n}$.
Therefore, due to the transfer of energy from plasmons localized in distant wires to the break,
a bath of bosonic excitations at the break is formed that allows to match the energy $E_a$
with the higher accuracy the more localized plasmons are involved in the creation of
a bosonic excitation.

In the disordered array the number of wires that can
supply energy to the break is limited by the plasmon localization length that
can be evaluated as $\xi_p=u_{g}\tau_p$, where
$\tau_p$ denotes the plasmon mean free time and $u_{g}$ is the plasmon group velocity.
The mean free time $\tau_p$ is related to the fluctuations of the
matrix element of the interwire charge density interactions
$\langle(\delta V_{i,i+1})^2\rangle=\frac{v_0}{(2\pi)^3\omega\nu_p\tau_p d}$,
where $\nu_p=\frac{L}{\pi^2v_1 n}$ is the plasmon density of states in the middle of the band
$n$, and $\omega$ is the plasmon frequency.
Furthermore, the random height of interwire tunneling barriers affects the interwire charge density
interactions
$V_{i,i+1}\sim e^2a^2|\psi_i|^2|\psi_{i+1}|^2/d$ through the randomness of
the wave functions $\psi_i$. Another source of randomness in the model is a random
charging energy due to the random environment (local concentration of charged impurities)
around each wire.
Random energy barriers are directly felt only by the exponentially decaying tail of the
single particle wave function within the barrier, whereas random fluctuations of the charge
density interactions are mostly determined by the part of the single particle wave function
within the wire. That is the reason why the disorder that leads to the localization of single particle
wave functions is felt only weakly  by the charge density fluctuations.
Using the normalization condition $|\psi_i|^2\left(a+\int_0^\infty e^{-2\lambda_i x}dx\right)=1$
one can relate the fluctuations of the decay factor $\lambda_i$
with the fluctuations of $|\psi_i|^2$ that in turn determines the fluctuations of the charge
density interactions and the plasmon mean free time. One obtains
$\tau_p=\frac{d(a\overline{\lambda}+1)^4v_0^2}{16\pi^2\omega^2 e^2a^2\langle\delta\lambda^2\rangle}$.
In the regime, when electrons are localized on the length $d$, we obtain
$
l_p/d=v_0^2(a\overline{\lambda}+1)^4e^{2\overline{\lambda}d}/
\left(2\omega\overline{\lambda}d\right)^2.
$
For $\overline{\lambda}d \gg \log\left(\frac{2\omega\overline{\lambda}d}{v_0(a\overline{\lambda}
+1)^2}\right)$, the plasmon localization length greatly exceeds the single particle localization
length.
If the length of array is less than $l_p$, the plasmons propagate freely along the array,
whereas the electrons are strongly localized. In that regime, the dispersion (\ref{u_p}) can
be used to describe the propagation of plasmons.

Treating the interwire charge density interaction perturbatively, we can write the
transition rate with the absorption of a plasmon similarly to a transition with the absorption
of a phonon using the Fermi golden rule \cite{Efros}
\begin{equation}
\gamma\propto \int dp \sum_{n}\sum_{m,k=0}^{\infty}|V_n(p)|N_B(\epsilon_{n}(p))
f\left(-E_m\right)
\left[1-f\left(E_a+E_k\right)
\right]\delta\left(\epsilon_{n}(p)-E_a-E_m-E_k\right).
\label{Golden-Rule}
\end{equation}
Here $V_n(p)$ is the strength of  interwire charge density interaction for the plasmon mode
$n$, $N_B(\epsilon_{n}(p))$ is the occupation number of the  plasmon mode,
$f(E_m)$ denotes the Fermi distribution and describes the occupation of the
$m$-th single-particle energy level in the wire, $E_m=\frac{\pi v_0}{L}m$.
For narrow plasmon energy bands, the perturbative approach suggests that
if  the energy $E_a$ lies in the gap between the plasmon bands the
hopping over the break is blocked.  This suggestion turns out to be wrong because
of a  conceptual difference between the plasmon and phonon transport mechanisms. Whereas the
phonons represent a bath of bosonic excitations that is independent of electrons,
the plasmons are ``made'' of electrons themselves.
Consequently, while the electron-phonon interaction can generally be
treated perturbatively, the perturbative treatment of plasmons is possible only under special
conditions. The applicability of the perturbative treatment of plasmons is determined by the
relation of two time scales: the characteristic time of plasmon dynamics $t_p$ and the
characteristic time of a single electronic hop $t_h$.
If the Coulomb interaction in a grain is well-screened or the plasmons are strongly
localized, then $t_p$ is the characteristic relaxation time of a plasmon excitation
within a single grain. For  $t_p\ll t_h$ the plasmons can be neglected
in transport. The description of interactions thus reduces to the capacitive model
\cite{AGK}.
For the delocalized undamped plasmons, the time $t_p$ is
associated with the formation of an extended in space plasmonic excitation. In that case,
$t_p\ll t_h$ correspond to the regime of a strongly nonlinear coupling between plasmons
and electrons, and the perturbative treatment of plasmons is incorrect.
Plasmons in one-dimensional wires represent a profound example for that regime.
For the model considered in this paper $t_p\sim L/v_1$, and at the break the condition $t_p\ll t_h$
is always satisfied. As we show below, due to the strong
electron-plasmon coupling, the nonlinear
effects lead to the creation of plasmon complexes with energies covering  the
whole spectrum continuously, even though the plasmon bands initially are very narrow.
This in turn leads to plasmon assisted transport with a temperature dependence that is
qualitatively different from the case of phonon assisted transport.
In the regime  $t_p\gg t_h$, the effective
interaction time is limited by $t_h$. Then the plasmon dynamics is essentially
independent of the electron dynamics and, in the case of a continuous plasmon spectrum,
the plasmon assisted transport is quite analogous to the phonon assisted one.

A unique feature of the chosen model is the applicability of the bosonized
description  that allows exact treatment of interactions and hence nonperturbative
treatment of plasmons. Precisely, the
plasmon dynamics is described by the action
\begin{equation}
S=\int_{-1}^{1}dp\int_0^\beta d\tau \int_{-L/2}^{L/2} \frac{dx}{2K_p}\left\{
\frac{1}{u_p}|\partial_{\tau} \Theta_p|^2+u_p|\partial_x\Theta_p|^2\right\},
\label{Sp}
\end{equation}
representing a finite size generalization of the sliding Luttinger liquid
model \cite{SLL}. The relation of the plasmon velocity $u_p$ and the
Luttinger liquid constant $K_p$ with inter- and intrawire interactions has
been calculated  in \cite{SLL}.
A fermion annihilation operator in the wire $n$, $\hat{\Psi}_n(x)$, is represented as
\begin{equation}
\hat{\Psi}_{n}^{\chi}(x)\sim
\hat{F}_{n}^{\chi}
\exp\left[-i\int_{-1}^{1}dp \phi_{p}^{\chi}(x) e^{-i\pi pn}\right],
\label{Psi-n}
\end{equation}
where $\chi=R,L$ denotes the chirality, $\phi_{p}^{\chi}(x)$ is a chiral bosonic field,
and $\hat{F}_{n}^{\chi}$ is a Klein factor.  The chiral field $\phi_p^{\chi}$ is in turn
expressed through the field $\Theta_p(x)$ and its dual $\Phi_p(x)$,
$
\phi_p^{R,L}(x)=\left(\Theta_p(x)\pm\Phi_p(x)\right)\sqrt{\pi}
$.

The resistance of the array is calculated along the lines of
Ref. \cite{RR}.
Let us parameterize the tunneling matrix element between two wires in the form
$
t_{i,i+1}=\exp(-|y_{i,i+1}|/d)
$.
The parameter $y$ can be associated with an effective distance between the two wires.
This effective distance is random, its distribution follows from the distribution of the
heights of potential barriers.
Since a break, being a junction with exponentially large resistance, is not
shorted by other resistances connected in parallel, we can write the resistance of a break in
the form
$
R_{1}=R_0\exp[2|y_{i,i+1}|/d+f(E_a,T)].
$
Here $E_a$ denotes an addition energy to transfer an electron over the break.
We remind that the disorder enters the model only as a random distribution of addition
energies $E_a$ and effective distances $y_{i,i+1}$.
The function $f(E_a,T)$ accounts for the effect of thermally activated
plasmons in the resistance of the break.
According to Ref. \cite{RR},
the probability density $\rho(u)$ for the resistance  $R/R_0=e^u$ is proportional to
$e^{-gA}$, where $A$ is the area in the $(y,E_a)$ phase
space that results in the resistance $e^u$, and $g$ is the linear
density of localized single particle states.
The resistance is calculated as
$
R=R_0 l_y\int_0^\infty du \  e^{u-gA(u)},
$
where $l_y$ is the length of the array.
Therefore, in order to calculate the resistance of the array in the localized regime,
we have to obtain an expression for the resistance of the break $R_1$. Since the break is not
shorted by other resistances, we conclude $R_1=1/\sigma_1$, where $\sigma_1$ is the conductance
of a break.
Assume that the break is formed by a junction between the wires with numbers
$0$ and $1$. We take the position of a pinhole connecting the
two wires as $x$.
In the linear response approximation the current through the break $I_1$ is determined
by  the correlation function \cite{Mahan}
\begin{equation}
X(\tau)=|t_{01}|^2\left\langle T_{\tau}\left(\Psi_0(x,\tau)\Psi^{\dagger}_1(x,\tau)\Psi_1(x,0)
\Psi^{\dagger}_0(x,0)\right)
\right\rangle
\label{X}
\end{equation}
that characterizes the probability of a single hop over the break,
$X_+(\tau)=X(\tau>0)$ and $X_-(\tau)=X(\tau<0)$.
Here $t_{01}$ is the tunneling matrix
element.
In the bosonized representation (\ref{Psi-n}),
the correlation function $X(\tau)$  factorizes in the correlator of Klein factors and the
correlator of bosonic exponents that we denote as $X_b(\tau)$.
The time dependence of the Klein factors $F_{n,\chi}(\tau)$ is given
by the ground state energy of the wire $n$ that includes the capacitive interaction between the
wires.  Thus the correlator of the Klein factors is proportional to $e^{-E_a\tau}$.
Furthermore,  calculating the correlators of bosonic exponents,
we cast the expression for the current into the form
$
I(eV)\propto \int_{-\infty}^{\infty}d\omega' J_0(eV+\omega')Z(-\omega').
\label{I-om-Z}
$
Here $J_0(\omega)\sim \delta(\omega-E_a)-\delta(\omega+E_a)$ defines the zero-temperature
current, whereas the influence of
thermally activated plasmons is contained in the factor $Z(\omega)$ that is given by the
the Fourier transform of
\begin{equation}
Z(t)=\exp\left[-\left\langle\kappa_p\sum_{\sigma=\pm 1}\sum_{m=1}^{\infty}
\ln\left(1- e^{-\frac{\pi u_p}{L}(m\beta+i\sigma t)}\right)\right\rangle_p
\right].
\label{Z(t)wd}
\end{equation}
Here $\kappa_p=K_p+1/K_p$, and the average over the plasmon wave vector $p$ is defined by
$\langle \cdot \rangle_p\equiv \int_0^{1}\cdot \left(1-2\cos(\pi p)+\cos(2\pi p)\right)dp$.
Further we assume the coupling constant  to be $p$-independent,
$\kappa_p=\kappa$ and use the simplified dispersion law (\ref{u_p}).
Note, that in approximate evaluations it is much more important to keep the
$p$-dependence of the velocity $u_p$ that reflects the formation of plasmon bands
than the $p$-dependence of the coupling constant $\kappa_p$. To the lowest order in
$p$-dependent terms, the latter just leads to the averaging of the single Luttinger liquid
result over the coupling constant.
The basic average to be used in subsequent calculations reads
\begin{equation}
\langle e^{-bu_p}\rangle_p=e^{-bv_0}\left[I_0(b v_1)+(1-\frac{1}{bv_1})
I_1(bv_1)\right],
\label{av-dp-I}
\end{equation}
where $b=(m\beta\pm it)\pi/L$, and
$I_{\nu}(z)$ denotes the Bessel function of complex argument.
For large times $t$, (\ref{av-dp-I}) gives asymptotically
\begin{eqnarray}
\langle e^{-\frac{\pi u_p}{L}(m\beta\pm it)}\rangle_p\approx
\sqrt{\frac{2L}{\pi^2|v_1|(m\beta\pm it)}}e^{\left[-\frac{\pi}{L}w(m\beta\pm it)\right]},
\label{av-p} &&
\end{eqnarray}
where $w=v_0-v_1$. Despite being obtained for $v_1<v_0$, (\ref{av-p}) is essentially
nonperturbative in $v_1$. The relevant values of the transport time are restricted by the
hopping time $\tau_h$. For $v_1|m\beta\pm i\tau_h|<1$ the correlations giving rise to
plasmon bands do not develop,  and the short time expansion of (\ref{av-dp-I}) has to be used
instead of (\ref{av-p}).
The latter is equivalent to the perturbative treatment of plasmons, leading to
a result similar to the phonon mechanism of hopping.

At low temperatures, $T\ll\frac{\pi w}{L}$,   the major contribution to $Z(t)$ in (\ref{Z(t)wd})
is given by the term with $m=1$.
Leaving only that term, expanding the logarithms, and substituting (\ref{av-p}) in
(\ref{Z(t)wd}) we obtain
\begin{equation}
Z(\omega)\approx\sum_{n,l=0}^{\infty}\frac{2\kappa^{l+n}}{n!l!}
\left(\frac{2L}{\pi^2|v_1|}\right)^{\frac{n+l}{2}}
e^{-\beta\left(\frac{\pi w}{L}(n+l)+|\omega-\epsilon_{ln}|\right)}
\left|\omega-\epsilon_{ln}\right|^{\frac{\nu}{2}-1}
\frac{\sin\left(\frac{\pi\nu}{2}\right)\Gamma\left(1-\frac{\nu}{2}\right)}{\left(2\beta
\right)^{\frac{l+n-\nu}{2}}},
\label{Z1}
\end{equation}
where $\epsilon_{ln}(\omega)=\frac{\pi w}{L}(l-n)$, and $\nu=l,n$ for
$\omega-\epsilon_{ln}\gtrless 0$ respectively.
Each term in (\ref{Z1}) describes a thermal excitation of a multiparticle plasmon complex with
a continuous density of states.
The plasmon complexes
described mathematically by (\ref{Z1}) form the bath of bosonic excitations that facilitate
transport over the break.
Higher values of $m$ in (\ref{Z(t)wd}) would correspond to excitations involving progressively
more plasmon modes. The leading terms in (\ref{Z1}) at low temperatures are given by
$l,n=\overline{0,1}$.
Those terms result in the leading low-temperature contribution to $\sigma_1$ in the form
\begin{equation}
\sigma_1\approx\frac{e^2}{\hbar}\frac{\kappa^2L\Gamma\left(\frac{1}{2}\right)}{
\sqrt{2}\pi^2|v_1|T}\left(|\beta E_a|^{-\frac{3}{2}}
+2|\beta E_a|^{-\frac{1}{2}}\right)
e^{-\beta\left(E_a+\frac{2\pi}{L}w\right)}.
\label{si1-approx}
\end{equation}
Further calculations closely follow Ref. \cite{RR} leading to
the temperature dependence of the resistance
\begin{equation}
R\approx \frac{\hbar}{e^2}
\frac{\pi^2|v_1| l_y T^{2}}{\sqrt{2}\kappa^2L\Gamma\left(\frac{1}{2}\right)} \exp\left[\left(\frac{1}{2gdT}+\frac{\pi w}{LT}\right)\right].
\label{R(T)}
\end{equation}
Since in the low-temperature regime the temperature broadening of discrete plasmon levels in a single wire is less than the interlevel separation, the finiteness of the wires essentially determines the physics of
transport. In particular, the LL interaction parameter $\kappa$ does not enter the temperature dependence
of the resistance, except as a general prefactor in (\ref{R(T)}).
For comparison, in the case of phonon-assisted transport in the same model, as well as
by a perturbative treatment of the plasmon-electron interactions the preexponential
factor in (\ref{R(T)}) goes like $T^{1/2}$. The result (\ref{R(T)}) also differs from the
thermally activated resistance of a single disordered wire obtained recently in Ref.
\cite{Nattermann}.

At high temperatures, when $T\gg\frac{\pi w}{L}$, the temperature broadening exceeds the
interlevel separation in a single wire. Then the plasmons in the array behave similarly
to plasmons in an array of infinite length wires. The conductance of a single junction depends on the addition energy not exponentially but as a power law $\sigma_1\sim E_a\cdot\left[\max\left(T, E_a\right)\right]^{\kappa-3}$. Therefore, the resistance of the array is not determined by the resistance of a single break, it is rather given by the average over the resistances of all junctions. The fact that the leading contribution to the conductance at high temperatures is given by the result for infinitely long wires implies that the coherence of the single particle motion is broken already by a single hop between the two neighbor wires. The latter justifies the averaging over the all junctions in calculation of the resistance. The expression for the resistance can
be written as a Drude formula with the interaction and temperature dependent mean free time
$\tau_f$. The expression for the mean free time
can be organized as an expansion in powers of the small parameter
$e^{-\frac{4\pi LT}{w}}$. At a typical charging energy $\bar{E}_a>T$, the
leading term in the expression for the mean free time is temperature independent,
$\tau_f\propto 1/\langle E_a^{2-\kappa}\rangle$. For comparison, phonon assisted transport in
that temperature regime still has a thermally activated character with the preexponential
factor $T^{-1/2}$ in (\ref{R(T)}).
At temperatures even larger than the typical charging energy,
$\bar{E}_a<T$, $\tau_f$ exhibits the power-low
temperature dependence typical for transport across  a sliding Luttinger liquid,
$\tau_f\propto T^{3-\kappa}/\langle E_a\rangle$ \cite{SLL}. Therefore, at high temperatures,
the mean free time is determined by the Luttinger liquid interaction parameter $\kappa$.
The calculation of the phonon assisted resistance in that regime using the Fermi golden rule
leads to the  Drude formula with the logarithmic temperature dependence of the mean free time $\tau_f\propto\ln T$.

In conclusion, we  demonstrated the possibility of plasmon assisted inelastic
transport in the particular case of a disordered granular array, an array of parallel
quantum wires. Due to large charge relaxation time in a wire,
the plasmon-electron interaction has to be treated nonperturbatively. In the result,
thermally activated resistance has qualitatively different temperature dependence
for plasmon assisted transport as compared to phonon assisted transport.
Despite the specific quasi one-dimensional geometry of the grains in this model,
the described phenomenon is believed to be of general importance for granular arrays with
delocalized or weakly localized plasmons.

\acknowledgments
The author is grateful to M. Raikh, who initiated this work, for numerous illuminating
discussions. The author appreciates fruitful discussions with D. Pfannkuche and
valuable comments of I. Gornyi and S. Kettemann.

\end{document}